\documentclass[aps,pra,twocolumn,groupedaddress]{revtex4-1}

\usepackage{graphicx}

\begin{document}


\title{Microstructured Stark Decelerator}


\author{Demitri Y. Balabanov}
\email[d\textunderscore balaba@uncg.edu]{}
\affiliation{Department of Nanoscience, Joint School of Nanoscience \& Nanoengineering, University of North Carolina, Greensboro, North Carolina, 27401, USA}

\author{Liam M. Duffy}
\email[liam\textunderscore duffy@uncg.edu]{}
\affiliation{Department of Chemistry \& Biochemistry, University of North Carolina, Greensboro, North Carolina, 27402-6170, USA}


\date{\today}

\begin{abstract}
A Stark decelerator is an effective tool for controlling motional degrees of freedom of polar molecules. Due to technical limitations, many of the current Stark decelerators focus on molecules in low-field-seeking quantum states and are built based on a fixed electrode size and spacing (type-A architecture). Here, we present an alternative method based on a different architecture, so-called type-B, with microstructured electrodes and a simpler electric field pulse timing with the prospect of producing cold, quasi-CW molecular beams. We demonstrate with simulations the feasibility of this method and show that a 2 cm decelerator consisting of 100 stages can bring molecules from 200 m/s to a near standstill in about 150 microseconds. Subsequently, the molecules can be reaccelerated to 200 m/s in a phase-stable manner. Our simulations focus on the longitudinal transmission leaving the problem of a transverse confinement for a future study. This work points to a possible way of integrating polar molecule accelerators with solid-state electronics. 

\end{abstract}


\maketitle

\section{Introduction}

The majority of the experimental efforts in Stark deceleration have focused on polar molecules in low-field seeking (LFS) quantum states. This is due to a number of technical challenges associated with efficient guiding, trapping and manipulation of molecules in high-field seeking (HFS) quantum states. From a scientific and a technological point of view, molecules in HFS states are a desirable system to work with for two major reasons. First, the ground state of any molecule is always high-field seeking with respect to an external perturbation. Ground states, in turn, are desirable due to their long lifetimes. Working with such a system could make it possible to create molecular Bose-Einstein condensates in traps \cite{Baranov2012}. This does not seem to be possible with LFS states due to undesirable collisions \cite{Bohn2001}. Second, the quantum states of large molecules become HFS in relatively small electric fields and all states become HFS given a high enough electric field. If it was possible to efficiently bring such a system to a cold (ultracold) regime, while retaining the control over the motional degrees of freedom, then the system could be a viable platform for quantum information processing \cite{Wei2011a}\cite{Wei2011}\cite{DeMille2002}. 

In the past several years, a number of microstructured devices have been proposed for manipulating polar molecules. They include a microstructured trap for storage and adiabatic cooling \cite{Englert2011}, a nano trap capable of confining microscopic and macroscopic polar particles \cite{Blumel2012}, a microstructured switchable mirror for reflection of a cold beam of state-selected polar molecules \cite{Schulz2004}, a microstructured elliptical mirror for focusing a beam of polar molecules \cite{Florez2011}, as well as a Stark decelerator on a chip \cite{Meek2009} and a guide used in conjunction with the sub-THz (mm-wave) radiation to address molecules located less than 50 $\mu m$ above the microstructured surface \cite{Santambrogio2011}. All of the above mentioned microstructured devices require voltages significantly above those used in solid-state electronics and work only on molecules in LFS quantum states. The integration of molecular systems, HFS molecular systems in particular, with solid-state devices remains challenging \cite{Carr2009}.

In this paper we present results on one-dimensional (1D) acceleration of the ground-state HCN-like molecular beam using a Monte-Carlo simulation method. The device can be used to slow down or to speed up molecular beams. Acceleration, both positive and negative, is achieved using an alternative type of Stark accelerator, the so-called \textit{type-B} \cite{Friedrich2004}, in which the electrode separation distance changes along the beam axis while the electric field switching time remains constant. This type of a device is in contrast to the commonly used \textit{type-A} in which the electrode separation distance is kept constant while the electric field switching time is varied. Section~\ref{s2} outlines the design of the microstructured \textit{type-B} Stark accelerator used in our experiments. We present the simulation results on acceleration of HCN-like molecular beam in section~\ref{s3}, followed by the summary and concluding remarks in section~\ref{s4}.

\section{Type-B Microstructured Stark Decelerator}\label{s2}
The advantages offered by the \textit{type-B} arrangement are the ability to produce cold, quasi-continuous wave (quasi-CW) molecular beams from CW molecular beams while using simpler electric field pulse timing. The miniaturization of the electrodes to the micrometer-scale enables the production of high electric field gradients using low voltages that are compatible with voltages used in microelectronics. The use of modern micro- and nanofabrication methods make possible the fabrication of electrodes with more precise alignment and tailored geometries. This is important because achieving precise electrode alignment, in particular, is the biggest technological hurdle limiting the efficiency of the current \textit{macroscopic} alternating-gradient (AG) decelerators for polar molecules \cite{Putzke2012}. Additionally, electrode shaping can maximize phase-space acceptance of the device and result in more favorable beam dynamics. Both the improved alignment and the electrode shaping would give greater control over the molecule's motional degrees of freedom.

In principle, one can devise arbitrarily complex electrode geometry to accelerate the molecular beam. Choosing a particular geometry determines the boundary conditions to the Laplace equation and therefore determines the electric field distribution. Knowing the spatial distribution of the electric field and the quantum state of the molecules inside the beam enables one to calculate numerically the force experienced by the molecules at any point inside the device. For simplicity, we chose an electrode geometry that does not change in one of the transverse directions (along x-axis). We neglect the edge effects due to fringe fields by extending all edges to infinity, making our electrode geometry effectively two-dimensional (2D). 

\subsection{Electrode Shape and Arrangement}
\begin{figure*}
\includegraphics[width=\linewidth]{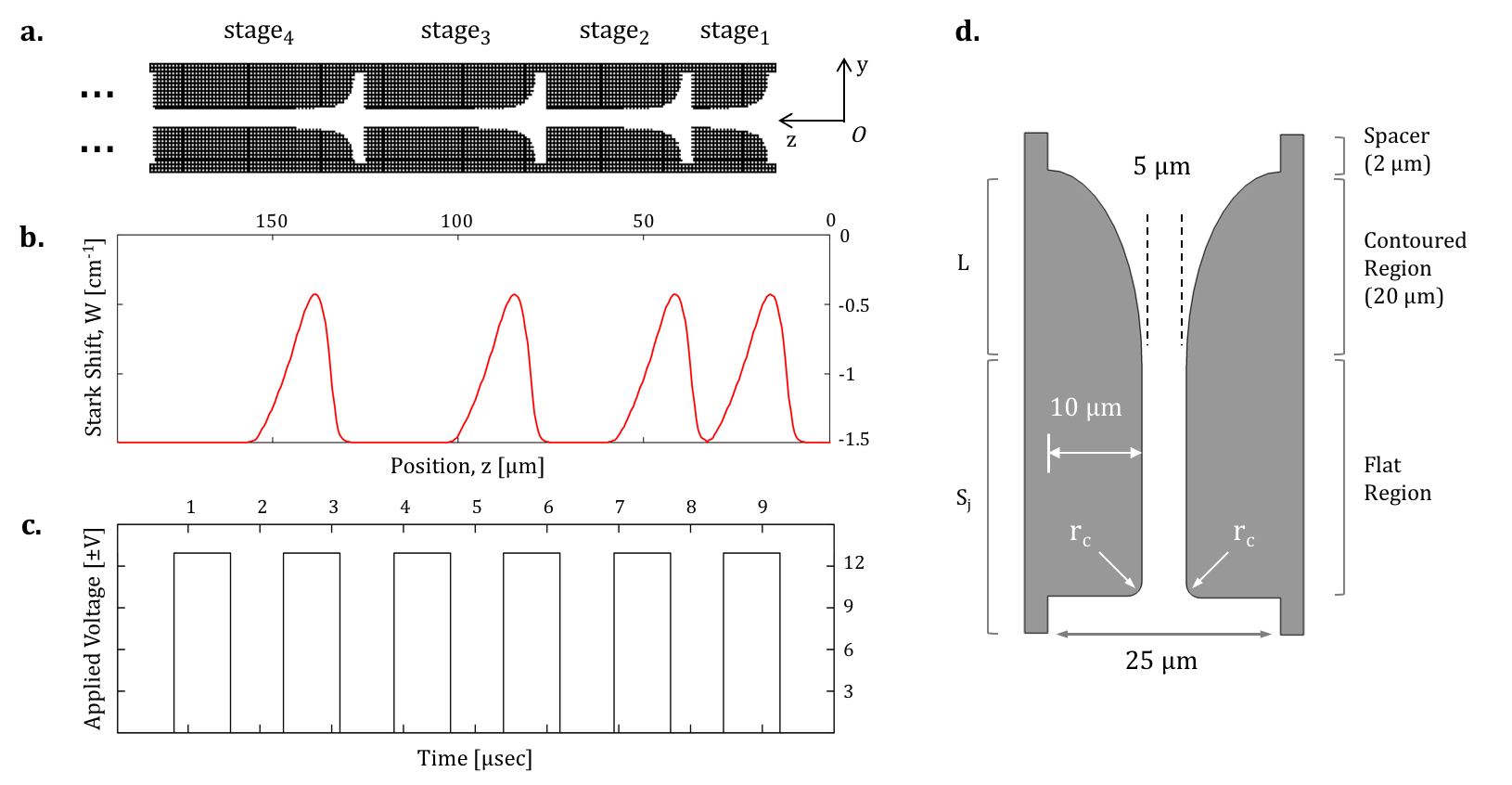}%
 \caption{\label{fig:fig1}Microstructured Type-B Accelerator-Decelerator for the Polar Molecules. (a) Layout of the \textit{type-B} accelerator-decelerator for polar molecules, showing the first four stages, in the case of acceleration, or the last 4 stages, in the case of deceleration. The lengths of the stages must either increase (acceleration) or decrease (deceleration), relative to the molecule's starting position, in order to have a constant electric field switching time. (b) Potential energy along the z-axis for HCN-like molecule in a ground ro-vibrational state (HFS). (c) The first 6 cycles of a constant-period, square-wave pulse sequence ($d = 0.511, T = 1535.4$ $nsec$) of $\pm 12.95$ $Volts$ applied to the decelerator. (d) Schematic view of a single stage. See text for details.}
 \end{figure*}

The shape and arrangement of the electrodes are illustrated in Figure~\ref{fig:fig1}. Two parallel, conductive plates are separated by a distance $l = 25$ $\mu m$. The surface of each plate is patterned with parallel, interconnected microstructures (Figure~\ref{fig:fig1}a). When a constant-period, square wave potential difference pulse is supplied to the plates (Figure~\ref{fig:fig1}c), these microstructures, which we call electrodes, produce regions of non-uniform electric field along the z-axis, the acceleration direction. To be consistent with the literature we define such regions of non-uniform electric field as field-stages \cite{Meerakker2008}. The microstructures act in a similar way as the electrodes in a conventional Stark or alternating-gradient (AG) decelerators except that we shape the electrodes to yield a particular longitudinal electric field distribution. For a molecule in a given quantum state, the exact shape of the electrodes and their arrangement determines the curvature profile of the resultant Stark (potential) energy hill and hence, determines the forces experienced by the molecule (Figure~\ref{fig:fig1}b). Note that, in principle, all electrodes could be insulated from each other and therefore be independently controlled. In particular, if the last stage is separate from all the other stages then a final (exit) beam velocity can be dialed by supplying that stage with an appropriate voltage.

Figure~\ref{fig:fig1}d shows a schematic of a single electrode stage. Each electrode is comprised of two regions: a fixed length contoured region and a variable length flat region. The shaping of the electrodes is accomplished by inputting a function inside the contoured region. One is free to choose any contour function. The contour function used in our experiments is a truncated right hyperbola which results in an approximately quadratic Stark energy profile. We've set the total length of the shortest stage to be 20 $\mu m$ and have limited the smallest feature size on the device to 1 $\mu m$ for ease of fabrication by lower-cost micro- and nanolithography techniques. A 5 $\mu m$ opening between the plates, through which the molecular beam propagates, was modeled recognizing that the requisite electric field strengths of 50\textendash 100 $kV/cm$ can then be achieved with conventional, commercially available, pulse generators operating at approximately $\pm$15 $Volts$. We note that higher separation distances can easily be used but would require supplying higher voltages. The stages are terminated with a 10 $\mu m$ perpendicular drop, which is also the width of the contoured region. Varying this width allows us to tune the electric field local minimum and therefore control the height of the potential energy hill seen by the molecules. All corners of the electrodes are rounded off with 1/4 circle of radius $r_{c} = 1$ $\mu m$ to reduce the undesired electric field gradients at the entrance (exit) of each stage.

\subsection{Spacing Between the Field-Stages}

\begin{figure}
\includegraphics[width=\linewidth]{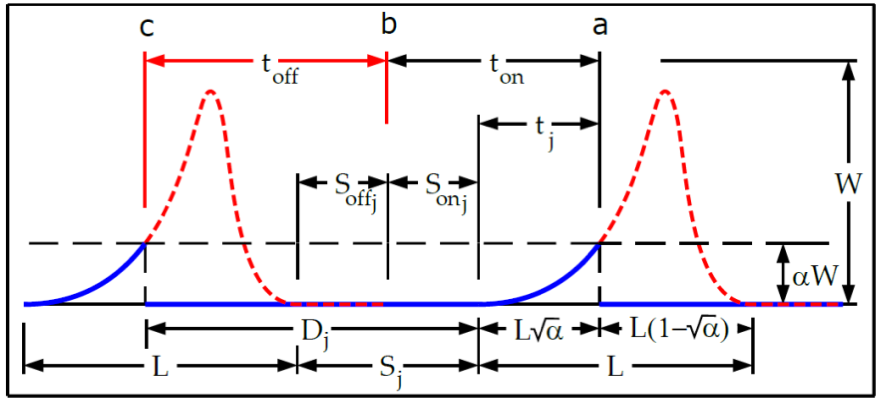}%
 \caption{\label{fig:fig2}Electrode separation, length, labeling and timing scheme. The longitudinal Stark potential (blue) experienced by a synchronous molecule traveling along the transverse center line between the plates. See text for details.}
 \end{figure}
 
Figure~\ref{fig:fig2} provides the field-stage separations, lengths, labeling and timing scheme. For the device presented in this article, all contoured regions of the plate surfaces are assumed to have a uniform shape and result in longitudinal Stark potentials of constant effective length $L$. Only the longitudinal separation (spacing) $S_{j}$ between the field-stages is changed to ensure that a constant duty cycle $d=t_{on}/(t_{on}+t_{off})$ can be used. This greatly simplifies the timing and device electronics since only two electrodes are needed and it also allows the device to operate on CW molecular beams, resulting in quasi-CW slowed beams. We use the labels of $L$ and $S$ to be consistent with the labeling used in the Alternating Gradient (AG) decelerator \cite{Bethlem2006a}. 

A synchronous molecule traveling along the transverse center line between the plates will experience a Stark potential like the one depicted by the thick solid (blue) line in Figure~\ref{fig:fig2}. In describing the operation of the device it is convenient to look at how the device would work as an accelerator. In such a mode, position (a) in Figure~\ref{fig:fig2} would correspond to the location of a stationary synchronous molecule waiting to be accelerated to the left (in the deceleration mode this is the final spot the synchronous molecules come to rest, after coming from the right). This also corresponds to the first contoured region of the device which we index as $j = 1$, with increasing indices to the left. If the plate voltages are switched ON at the time and position corresponding to the point (a), a stationary synchronous molecule at this position will be accelerate to the left over an acceleration time corresponding to $t_{j=1}$ and then travel at a constant velocity $v_{1} = \sqrt{2\alpha W/M}$ to position (c), where $M$ is the mass of the molecule, and $\alpha$ is the fraction of the maximum Stark energy $W$ added (or removed) from a synchronous molecule in each contoured field region. This constant velocity region is denoted by length $D_{1}$. Note that along the way, at point (b), the field is turned OFF so that this synchronous molecule never experiences the other parts of the Stark potential depicted by the (red) dashed line. A synchronous molecule will arrive at point (c) just as the field is turned back ON.

An average force approximation for determining pulse timings is frequently described in Stark-AG decelerator literature \cite{VandeMeerakker2012}. While this approximation is reasonable at higher velocities and large electrode separations, for the device described here, the variation in acceleration time $t_{j}$ with field region $j$, can have a significant impact on designed $S_{j}$ lengths between contoured features. To address this problem, we instead assume a harmonic form for the longitudinal Stark potential. We calculate the appropriate $S_{j}$ spacing as $S_{j} = D_{j} - L(1-\sqrt{\alpha})$, where $D_{j} = v_{j}(t_{on}+t_{off}-t_{j}) = v_{j}(t_{on}/d - t_{j})$, $v_{j}$ is the longitudinal velocity of a synchronous molecule in the constant velocity region between the contoured regions, $d$ is the duty cycle as defined above, and $t_{j}$ is the time spent in each acceleration (deceleration) region. In turn, $t_{on}$ is determined from $t_{on} = t_{1}+S_{1}\beta/v_{1}$, where $\beta$ is the fraction of the way along $S_{1}$ that the molecule travels before the field is turned OFF.

If we assume a harmonic form for the longitudinal Stark potential, then it can be shown that the time a molecule spends in each acceleration region depends on its velocity according to the relation

\begin{equation} \label{eq:eq1}
t_{j}=\frac{1}{\omega}\sin^{-1}\left({\frac{\omega L \sqrt{\alpha}}{v_{j}}}\right)=\frac{L \sqrt{\alpha}}{v_{1}}\sin^{-1}\left({\frac{1}{\sqrt{j}}}\right)
\end{equation}
where $\omega$ is the harmonic angular frequency of the Stark potential and the second equality comes from noting that the molecule's velocity varies with $j$ as $v_{j} = v_{1}\sqrt{j}$ and that $\omega = \frac{v_{1}}{L\sqrt{\alpha}}$. Combining expressions, the longitudinal spacing between contoured regions then goes as

\begin{widetext}
\begin{equation} \label{eq:eq2}
S_{j} = S_{1}\sqrt{j}+L\sqrt{\alpha j }\left(\frac{\pi}{2}-\sin^{-1}\left(\frac{1}{\sqrt{j}}\right)\right)+L(1-\sqrt{\alpha})(\sqrt{j}-1)
\end{equation}
\end{widetext}
and the duty cycle becomes

\begin{equation} \label{eq:eq3}
d=\frac{\frac{\pi}{2}L\sqrt{\alpha}+S_{1}\beta}{S_{1}+\frac{\pi}{2}L\sqrt{\alpha}+L(1-\sqrt{\alpha})}
\end{equation}

A design goal of the current simulations was to make the device as short as possible, which occurs for $S_{1} = 0$, which simplifies these expressions to those used in the studies herein as

\begin{equation} \label{eq:eq4}
S_{j}=L\sqrt{\alpha j}\left(\frac{\pi}{2}-\sin^{-1}\left(\frac{1}{\sqrt{j}}\right)\right)+L(1-\sqrt{\alpha})(\sqrt{j}-1)
\end{equation}
and

\begin{equation} \label{eq:eq5}
d=\frac{\pi}{2}\left(\frac{\pi}{2}+\frac{1}{\sqrt{\alpha}}-1\right)^{-1}
\end{equation}

In this scheme, the position of the synchronous particle is given by 

\begin{equation} \label{eq:eq5b}
x_{j} = L(1-\sqrt \alpha) + \sum\limits_{k=1}^j (S_{j} + L)
\end{equation}

\section{Experimental Results}\label{sec:expRes}\label{s3}

The calculations were performed using the latest version of SIMION \cite{Manura}, a charged particle optics simulation software, which we have modified to work with polar molecules. The program uses a highly modified $4^{th}$ order Runge-Kutta method for numerical integration of the trajectories and over-relaxation finite difference technique for finding solutions to the Laplace equation. 

The parameters of the simulation particles were equated to the physical parameters of hydrogen cyanide (HCN). In particular, HCN is a simple linear tri-atomic molecule with a relatively large dipole moment of 2.98 $Debye$ in its ground state \cite{Tomasevich1970}, molar mass of 27.03 $g/mol$ and a rotational constant of 44315.975 $MHz/c$. HCN is also an attractive molecule for manipulation with mm-wave techniques. The simulated particles were assumed to be in a ground ro-vibrational state. 

To estimate the resultant Stark shift and the effective dipole moment we've used a high-field, pendular-state model \cite{Bethlem2006a}. In this model the Stark shift $W$ is

\begin{equation} \label{eq:eq6}
\frac{W(\nu_{p},\lambda)}{B}=-\lambda+(\nu_{p}+1)\sqrt{2\lambda}
\end{equation}
and the effective dipole moment may be expressed as
\begin{equation} \label{eq:eq7}
\mu_{eff}=\mu \left(1-\frac{\nu_{p}+1}{\sqrt{2\lambda}} \right)
\end{equation}
where $\lambda=\mu E/B$ is the dimensionless parameter relating the body-fixed dipole moment $\mu$ and the rotational constant $B$ of an idealized rigid-rotor molecule interacting with the electric field $E$. Application of the electric field readily mixes different angular momentum quantum states $J$. Therefore, in the strong-field limit $\lambda \gg 1$, the pendular states must be characterized by the quantum numbers $\nu_{p}$ and $M$ (projection of $J$ onto the field axis) with $\nu_{p} = 2J - |M|$. In this limit, the body-fixed dipole moment is essentially parallel to the applied electric field and all the low-lying states are HFS. Furthermore, for small changes in the electric field, the effective dipole moment can be approximated as constant. 

\subsection{Deceleration}

In general, molecular species require different deceleration voltages depending on their initial kinetic energy and Stark shift. Applying $\pm$12.95 $V$ to the electrodes would result in a maximum electric field of 51.78 $kV/cm$ along the beam axis. At such electric fields HCN-like moleculeÕs effective ground-state dipole moment is equal to 1.39 $Debye$ with the total high-field limit Stark shift of about $\Delta W_{max}=1.81$ $cm^{-1}$ (or 0.224 $meV$). In order to decelerate a bunch of molecules from an initial longitudinal speed of $v_{0}=200$ $m/s$ to a complete stop using $n = 100$ stages requires each field-stage to extract $\Delta E=\frac{1}{2}mv^2_{0}/n$ of moleculeÕs kinetic energy. For HCN-like particle this equals to 0.451 $cm^{-1}$ (at $\Delta E/\Delta W_{max}=\alpha=0.25)$. As the molecules propagate through the device we repeatedly turn the electric field ON for the duration of $t_{on}=785.4$ $nsec$, and then turn it OFF for the duration of $t_{off}=750.0$ $nsec$ so that the duty cycle is equal to 0.511.

\begin{figure}
\includegraphics[width=\linewidth]{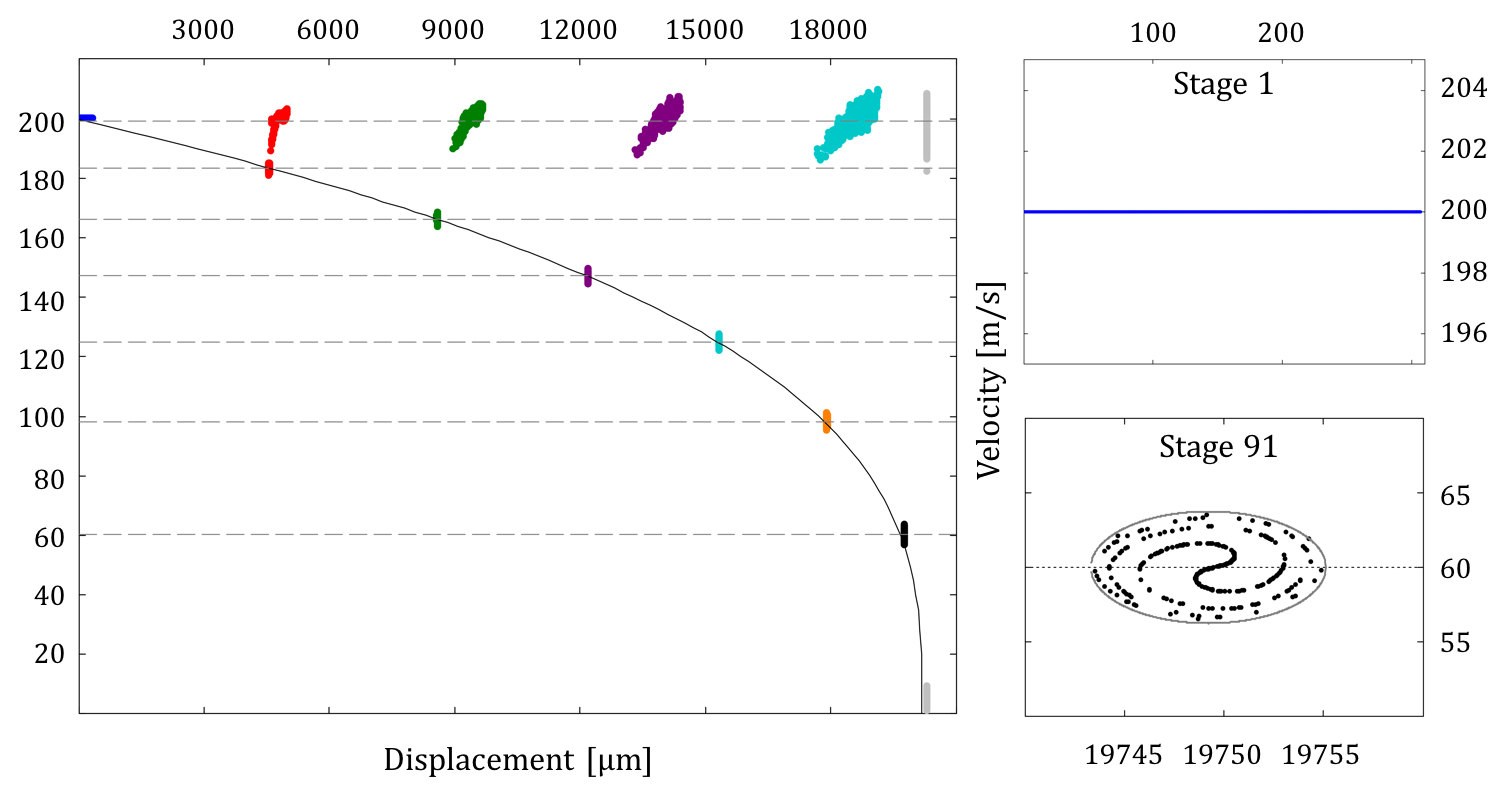}%
 \caption{\label{fig:fig4}Phase-space plot of the deceleration process for a ground state, HCN-like monochromatic molecular beam. A small subgroup of the initial distribution, shown in the insert, forms a bunch and is slowed down in a phase stable manner. Solid (black) curve is a theoretical prediction for the velocity and position of the synchronous molecule in the instantaneous impulse approximation. Horizontal (dashed) lines correspond to the expected velocities of the synchronous molecule at various stages. Data is recorded at stages 1 (blue), 16 (red), 31 (green), 46 (purple), 61 (teal), 76 (orange), 91 (black), 100 (grey). The last recording contains only the velocity information as it was recorded at a single point, along the z-axis, corresponding to the end of the decelerator.}
 \end{figure}

In order to confirm that the field-stages have the proper spacing, we inject a monochromatic beam of molecules and monitor the phase-space dynamics of the deceleration process. Initially, molecules are distributed uniformly along the beam axis, spanning the entire length of the first stage, with a longitudinal speed of 200 $m/s$ (see Figure~\ref{fig:fig4}, upper right inset). As the molecules move through the device a certain subset of the initial molecular distribution is gradually decelerated to about 5 $m/s$ after traversing 100 stages. Due to phase stability the molecules within this bunch are held together, maintaining a narrow spatial spread and velocity spread throughout the deceleration process. The lower right insets of Figure~\ref{fig:fig4} shows a phase-space bucket at stage-91 enclosed by an ellipse. Molecules within the ellipse execute elliptic, non-overlapping trajectories around the synchronous molecule located at the center. Molecules located outside of the ellipse are not amenable to a phase-stable deceleration and are considered ÔlostÕ, pulling further ahead of the decelerated bunch. It should be noted that our last recording, at the end of stage-100, is meant to simulate a molecular beam hitting a CCD plate detector, which is fixed at the end of the device. Hence, the last recording produces only the time and velocity data.

\begin{figure}
\includegraphics[width=\linewidth]{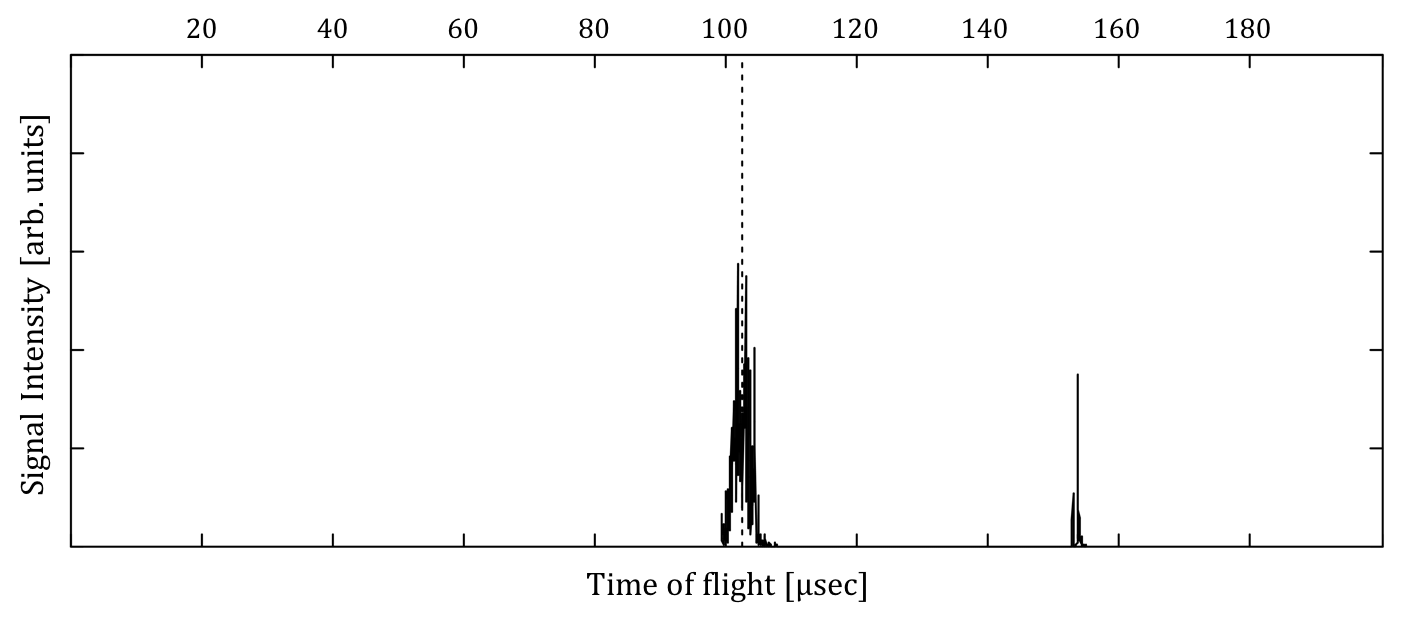}%
 \caption{\label{fig:fig5}Time-of-flight (TOF) distribution of the ground-state HCN-like molecular beam with an initially uniform spatial distribution and a starting velocity of 200 $m/s$. The second peak at around 153 $\mu sec$ corresponds to the arrival time of the decelerated bunch. The vertical dashed line indicates an expected arrival time of a molecule flying with a constant speed of 200 $m/s$.}
 \end{figure}

The acceptance of the decelerator corresponds to only a small fraction of the initial distribution (7 \%). Therefore, for most of the molecules the spacing of the electrodes and the field timing sequence will result in no significant (net) acceleration or deceleration. This molecular group, being largely unaffected by the presence of the deceleration stages, maintains the initial speed of 200 $m/s$ although with an increased velocity and spatial spread. This is clearly seen in the time-of-flight (TOF) distribution in Figure~\ref{fig:fig5}. A molecule flying with a constant speed of 200 $m/s$ would take about 102 microseconds to exit a 2.04 $cm$ long decelerator, which corresponds to the mean arrival time of the ÔunaffectedÕ molecular group (the first peak in Figure~\ref{fig:fig5}). The smaller, decelerated bunch arrives at the exit after about 153 microseconds (the second peak in Figure~\ref{fig:fig5}). 

Also plotted in Figure~\ref{fig:fig4} are the expected velocities assuming 0.451 $cm^{-1}$ reduction in the kinetic energy per stage (dashed lines) along with the theoretical predictions (solid line) for the speed and position of the synchronous particle using an instantaneous impulse approximation. In this approximation it is assumed that the synchronous particle receives an instantaneous reduction in its kinetic energy, equal to $\Delta E$, each time an electric field is turned ON. While the position for a synchronous particle in a harmonic potential is given by Equation~\ref{eq:eq5b}, the position of a synchronous particle under the instantaneous impulse approximation is given by
\begin{equation} \label{eq:eq8}
x_{j}=x_{0}+(t_{on}+t_{off})\sum\limits_{k=1}^j \sqrt{v^{2}_{0}\mp k\frac{2}{m} \Delta E}
\end{equation}
where $x_{0}$ is the starting position relative to the origin, $v_{0}$ is the initial speed of the synchronous particle whose mass is $m$. The first and last stages encountered by the molecule are $j=1$ and $j=100$, respectively. Positive sign is used in the case of acceleration and negative in the case of deceleration. Since there is no information regarding the shape of the potential hill in Equation~\ref{eq:eq8}, the above formula is general and may be used for various electrode geometries. 

It should be kept in mind, however, that this approximation tends to be less accurate as the electrode lengths decrease. The speed of the molecule decreases with each stage while the length of the contoured region remains constant. Therefore, the time over which the molecule experience a force gets longer as the electrode lengths decrease. The inaccuracies in the electrode spacing, due to this approximation, accumulate with an increase in field-stage number. Having an accurate (numerical) solution for the Stark energy hill from the chosen electrode geometry would allow one to properly account for the particleÕs transit time within each stage and obtain the required stage spacing.

\begin{figure}
\includegraphics[width=\linewidth]{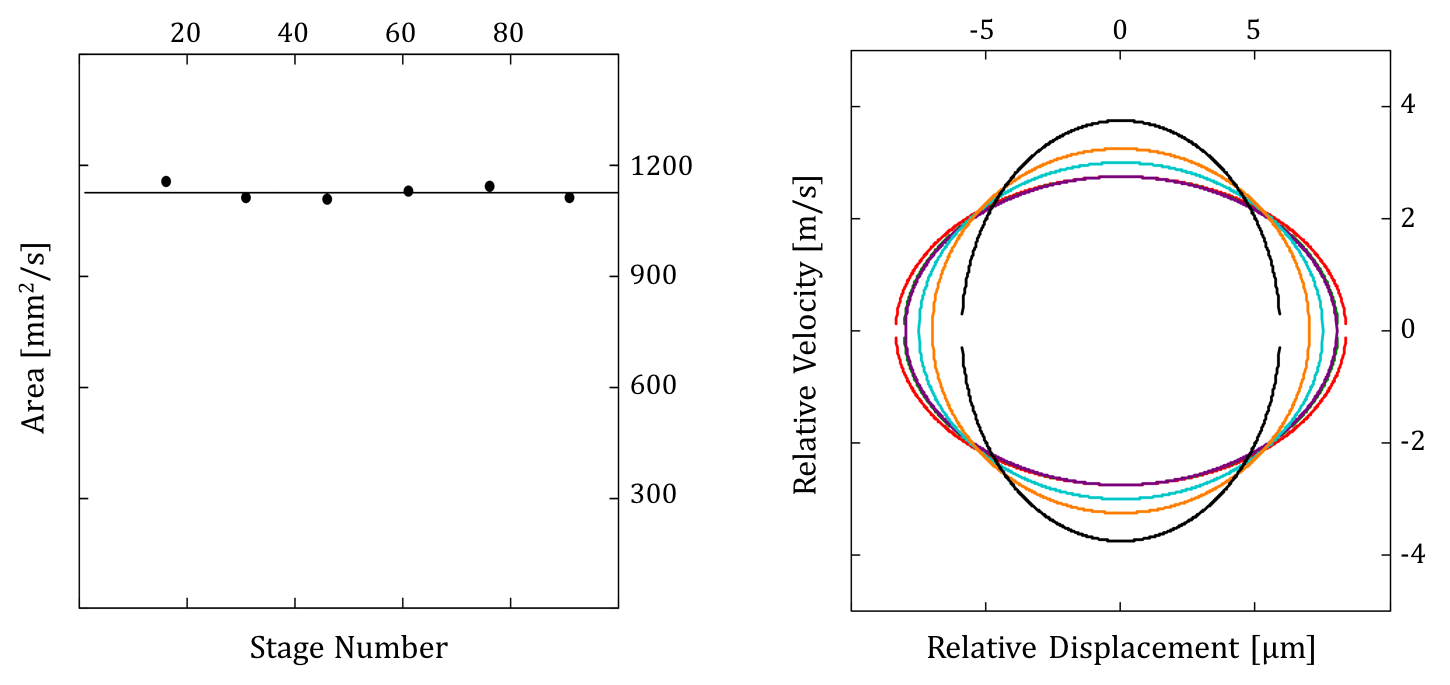}%
 \caption{\label{fig:fig6}(a) Measured phase-space area of the decelerated bunch. (b) Gradual reduction in the spatial spread and the concomitant increase in the velocity spread for stages 16 (red), 31 (green), 46 (purple), 61 (teal), 76 (orange) and 91 (black).}
 \end{figure}

As mentioned above, particles comprising the initial phase-space filament execute elliptical trajectories inside the bucket, delineated by the phase-space ellipse. Since only conservative forces are involved in the deceleration, the phase-space area must remain constant via Liouville theorem (Figure~\ref{fig:fig6}a). Hence, minimizing the spatial spread maximizes the velocity spread and vice versa. In the phase-space representation this can be seen as a 90$^{\circ}$ rotation of the phase-space distribution. In figure~\ref{fig:fig6}b, we plot the phase-space ellipses corresponding to stage 16 (red), 31 (green), 46 (purple), 61 (teal), 76 (orange) and 91 (black). Gradual reduction in the spatial spread and the concomitant increase in the velocity spread are clearly observed.

\begin{figure}
\includegraphics[width=\linewidth]{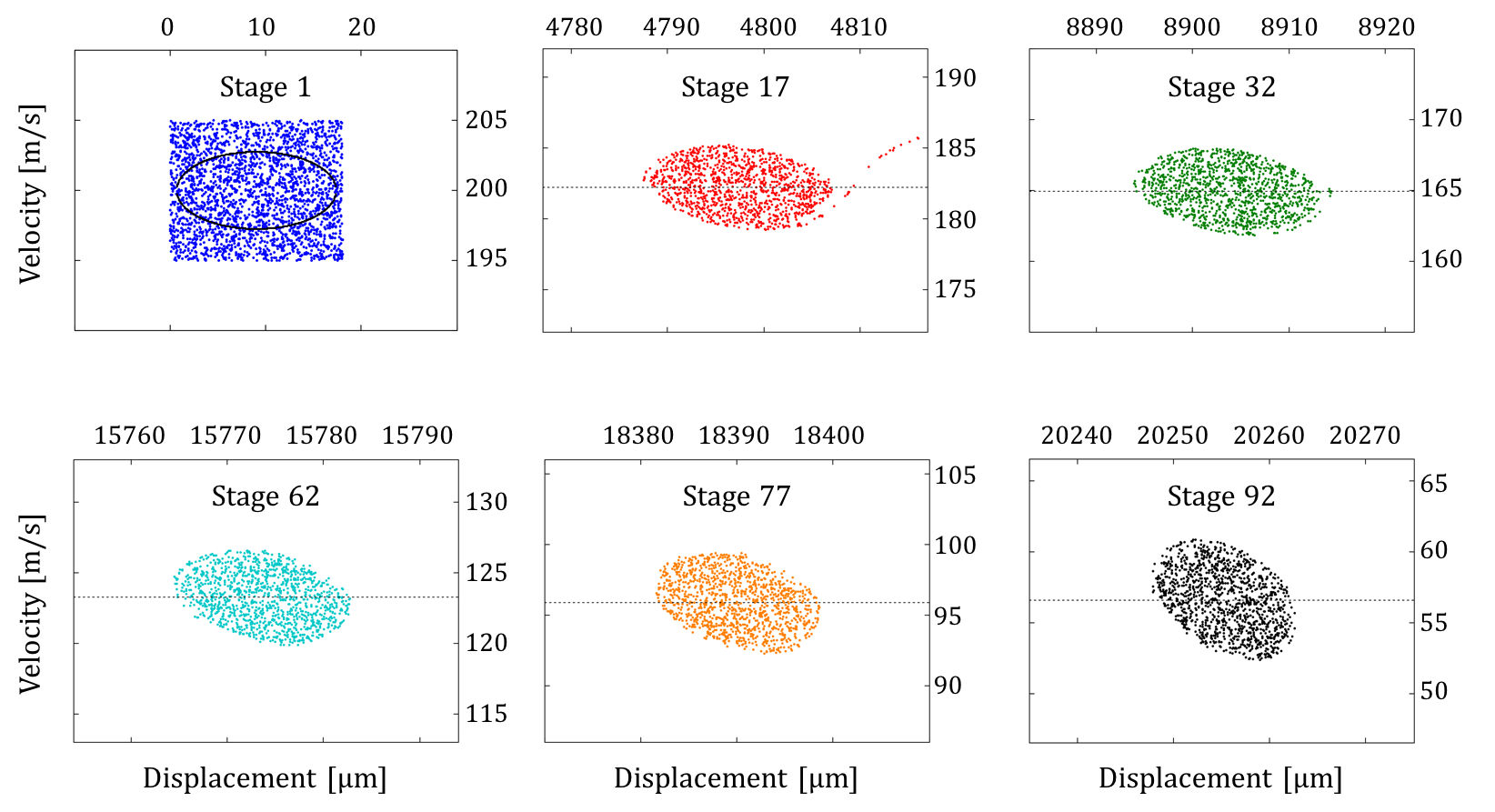}%
 \caption{\label{fig:fig7}Phase-space plot of the deceleration process for a ground state, HCN-like molecular beam with an initial uniform velocity distribution of $200 \pm 5$ $m/s$, recorded at stages 1, 17, 32, 62, 77, and 92.}
 \end{figure}

Another way of seeing this rotation is to simulate a molecular beam that has not only spatial distribution but also a uniform velocity distribution from 195 $m/s$ to 205 $m/s$ (see Figure~\ref{fig:fig7}). The plot for stage 1 shows that the phase-space acceptance of the device lies well within the initial distribution. Those molecules located outside of the acceptance region will eventually be lost from the decelerated bunch while resulting in a typical Ôgolf clubÕ phase-space distribution (Stage 17). The bucket itself rotates uniformly in phase-space and can be made to assume any desired angle. For example, a rotation of 90$^{\circ}$ results in a maximum narrowing of spatial spread of the decelerated bunch.

This focusing in the forward direction, or Òspatial bunchingÓ, of the molecular beam may be of an advantage for experiments where it is desirable to increase the number density at a given point along the beam axis. This would include collision studies as well as loading of the decelerated beam into an electrostatic trap \cite{Bethlem2002}. Alternatively, one can reduce the temperature of the decelerated beam by rotating the bucket such that the velocity spread is minimized. This would result in molecules keeping together for a longer period of time, albeit, at an increased spatial separation.

\subsection{Acceleration}

The topic of acceleration of polar molecules, in contrast to deceleration, has not been investigated much. Yet a velocity-controlled acceleration (or re-acceleration) of a beam (or a group of molecules) is a relevant process for manipulation of neutral molecules. It may, indeed, prove useful to accelerate the molecules before decelerating them or to have acceleration-deceleration cycles in order to achieve some desired beam (ensemble) property. As an example, acceleration of polar molecules in a \textit{type-B} accelerator gives rise to the phenomenon of phase compression, in the small-angle oscillation regime, which is absent in acceleration process of type-A device or in the deceleration of molecules in either \textit{type-A} or \textit{type-B} \cite{Friedrich2004}. Acceleration of neutral molecules may also make it possible to use non-conservative forces, such as radiation damping, to reduce the phase space volume occupied by the beam distribution, resulting in beam cooling. In 2004, Friedrich group published a paper on quasi-analytic model of a linear Stark accelerator-decelerator \cite{Friedrich2004} in which some of the dynamics underlying acceleration of polar molecules by \textit{type-A} and \textit{type-B} accelerator were examined.

\begin{figure}
\includegraphics[width=\linewidth]{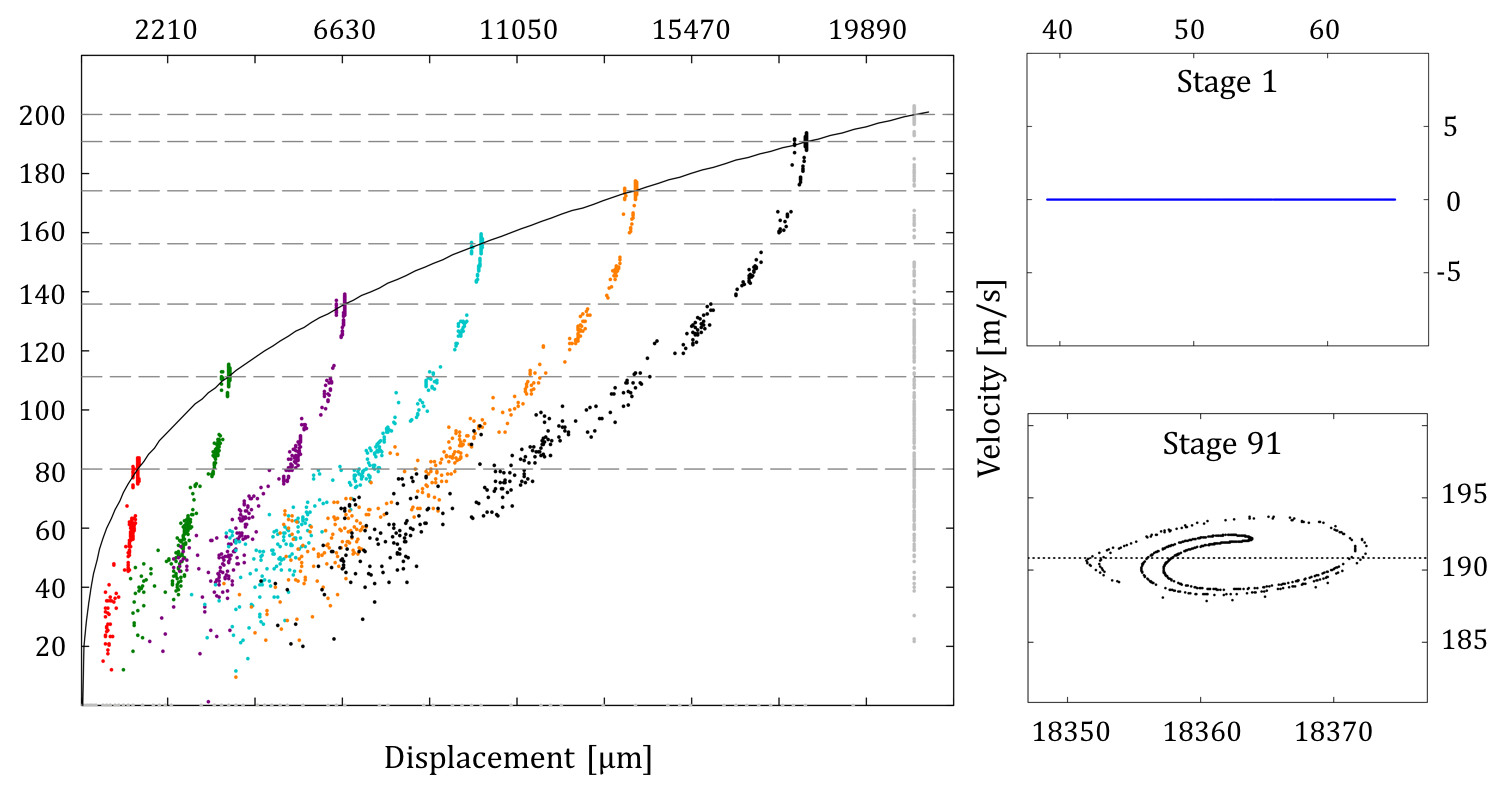}%
 \caption{\label{fig:fig8}(a) Phase-space plot of the acceleration process for a ground state, HCN-like molecular beam, recorded at stages 1, 16, 31, 46, 61, 76, 91 and 100 using the same color scheme as before. Also shown is the theoretical prediction for the velocity and position of the synchronous molecule in the instantaneous impulse approximation (solid curve). Horizontal (dashed) lines correspond to the expected velocities of the synchronous molecule at the recorded stages. (b) In this simulation all molecules have an initial velocity of 0 m/s and are spread uniformly, along the beam axis, over the entire length of stage-1. (c) Molecules comprising the leading, accelerated bunch maintain phase stability throughout the acceleration process.}
 \end{figure}

Here, we present simulation results on acceleration of polar molecules. The scheme is similar to the one used in our deceleration studies where only conservative forces were involved. The electrodes are activated for the duration of 785.4 $nsec$ with a duty cycle of $d = 0.5$. In order to accelerate molecules from standstill to 200 $m/s$ after passing through 100 stages requires each molecule to receive 0.451 $cm^{-1}$ of the kinetic energy per stage. Figure~\ref{fig:fig8}a shows a phase-space plot of the acceleration process for a ground state, HCN-like molecular beam recorded at every $15^{th}$ stage starting from stage-1. As before, the molecules slam into the detector positioned at the end of the device. Most of the molecules in the initial phase-space filament (Figure~\ref{fig:fig8}b) are contained in the fastest, leading bunch while the rest of the molecules acquire slower velocities and form groups within a Ôcomet-tail-likeÕ structure. The phase stability is maintained, for the leading bunch, throughout the acceleration process (Figure~\ref{fig:fig8}c). Using an instantaneous impulse approximation, it is possible to predict the position and speed of the synchronous particle. The solid (black) curve in Figure~\ref{fig:fig8}a is obtained using Equation~\ref{eq:eq8} and agrees with the simulation results to within $\pm2$ $\mu m$. 

\section{Conclusion}\label{s4}

Stark deceleration of polar molecules is a versatile method for producing slow and cold beams. Many of the devices thus far have been using \textit{type-A} architecture where the electrode spacing is kept constant. In this article we presented simulation studies on an alternative, \textit{type-B}, architecture where the electrode spacing changes in such a way so as to keep the electric field switching time constant, thus greatly simplifying the driving electronics. Building such a decelerator on a macro-scale ($\approx 1$ $m$) is much more challenging, compared to building a \textit{type-A} device, due to variable electrode spacing. However, miniaturization of the device can readily achieve the desired spacing and electrode alignment using standard micro- and nanolithography techniques while making it possible to operate the device at low voltages. It also becomes possible to shape the electrodes in order to maximize the phase-space acceptance.

We show that a 2 $cm$ decelerator consisting of 100 stages can bring molecules from 200 $m/s$ to a near standstill in about 150 microseconds. To decelerate faster beams it is possible to use longer decelerators, utilizing more stages, or higher voltages. However, in order to effectively decelerate molecules to low speeds requires having an accurate numerical solution for the potential energy hills given particular electrode geometry. Ultimately, the proper electrode geometry and stage spacing should be done by an optimization computer code, similar to how it is done in designing particle accelerators where the entire beam transport line is simulated.

The same device can be used to accelerate stationary or slow moving molecules loaded in traps. A phase-stable acceleration from 0 $m/s$ to 200 $m/s$ using 100 stages is possible. We suggest that the acceleration of molecules may be desirable for achieving certain beam properties, such as phase compression, before ultimately decelerating the beam. One can think of acceleration-deceleration cycles utilizing frictional forces that reduce velocity by radiation, resulting in beam cooling. 

There exist a number of methods for producing cold beams of various species of atoms and molecules. Two of the most popular ones are the supersonic expansion and the buffer gas cooling. Supersonic beam expansion results in a very narrow forward velocity spread and therefore produces some of the coldest beams ($\approx 100$ $mK$) which, due to enthalpy conservation, travel at supersonic velocities. Seeding can significantly reduce the forward velocity of such a beam to around 300 $m/s$. In contrast, buffer gas beams have a somewhat larger forward velocity spread but a lower average forward velocity around 150 $m/s$ \cite{Hutzler2012}. In our experiments, we used 200 $m/s$ as the initial velocity of the molecular beam in order simulate conditions that are within reach of both methods. 

The problem of transverse stability of molecular beams in macroscopic Stark decelerators has been addressed elsewhere \cite{Kalnins2002}\cite{VandeMeerakker2012}. A similar approach may be applied to studying transverse stability in microstructured decelerators like the one presented in this paper. We have identified a few promising electrode arrangements that should provide transverse stability to the molecular beams in microstructured decelerators \cite{Balabanov2015}. A full simulation study will be presented in a future publication. 

\begin{acknowledgments}
We would like to thank Shyam Aravamudhan and Jun Yan for useful discussions on microchip fabrication. 
\end{acknowledgments}

\bibliography{authorsbib.bib}

\end{document}